**Inelastic ponderomotive scattering of electrons at a high-intensity optical travelling wave in vacuum**

M. Kozák[1,2,*], T. Eckstein[1], N. Schönenberger[1], and P. Hommelhoff[1]

[1] Department of Physics, Friedrich-Alexander-Universität Erlangen-Nürnberg (FAU), Staudtstrasse 1, 91058 Erlangen, Germany, EU

[2] Faculty of Mathematics and Physics, Charles University in Prague, Ke Karlovu 3, 12116 Prague 2, Czech Republic, EU

* e-mail: martin.kozak@fau.de

**In the early days of quantum mechanics Kapitza and Dirac predicted that matter waves would scatter off the optical intensity grating formed by two counter-propagating light waves [1]. This interaction, driven by the ponderomotive potential of the optical standing wave, was both studied theoretically and demonstrated experimentally for atoms [2] and electrons [3-5]. In the original version of the experiment [1,5], only the transverse momentum of particles was varied, but their energy and longitudinal momentum remained unchanged after the interaction. Here, we report on the generalization of the Kapitza-Dirac effect. We demonstrate that the energy of sub-relativistic electrons is strongly modulated on the few-femtosecond time scale via the interaction with a travelling wave created in vacuum by two colliding laser pulses at different frequencies. This effect extends the possibilities of temporal control of freely propagating particles with coherent light and can serve the attosecond ballistic bunching of electrons [6], or for the acceleration of neutral atoms or molecules by light.**

Depending on the scattering regime, the interaction between electrons and the ponderomotive potential of an optical standing wave can be described both quantum mechanically (Kapitza-Dirac effect [7-9]) or classically [8,10]. In the quantum picture, the matter wave coherently diffracts on the periodic potential of the two colliding optical waves with wavevectors **k** and −**k** and identical frequency $\omega$, leading to observation of a series of diffraction peaks separated by two photon recoils $2\hbar\mathbf{k}$ (Raman-Nath regime [5,7]) or an individual diffraction peak (Bragg regime [8,11]). From the point of view of energy and momentum conservation, a diffracted particle simultaneously absorbs a photon from the first wave and emits a photon to the second wave via stimulated Compton scattering. The strength of the interaction is proportional to the light intensity (density of photons) of the optical standing wave.

From the classical perspective describing an incoherent scattering regime [4,8,10], the periodic ponderomotive potential of the optical standing wave $U_\mathrm{p} = e^2 |E_0|^2 / (m_0 \omega^2) \cos^2(\mathbf{k}.\mathbf{r})$, where $e$ is electron charge, $m_0$ is electron mass and $E_0$ is the field amplitude of each wave, leads to scattering of electrons due to the spatial dependence of the transverse ponderomotive force $F_\mathrm{p} = -\nabla U_\mathrm{p}$. Also more general cases of inelastic scattering of electrons by two-color fields were proposed [12-15], and ponderomotive ballistic bunching of electron beams was theoretically considered [6,16,17]. However, experiments along these lines have not been realized hitherto.

In this paper we experimentally demonstrate a strong modulation of energy and longitudinal momentum of electrons (momentum component in the direction of electron propagation) in vacuum by using two pulsed laser beams at different frequencies $\omega_1$ and $\omega_2$ intersecting with a pulsed electron beam at non-zero angles of incidence $\alpha$ and $\beta$ (Figure 1**a**-**c**). These laser fields create an optical travelling wave propagating parallel to the electron beam with a group velocity $v_\mathrm{g} = (\omega_1 - \omega_2) c / (\omega_1 \cos\alpha - \omega_2 \cos\beta)$, which can be synchronized with the electron initial velocity $v_\mathrm{i}$ (see Supplementary Information for details). The electrons thus inelastically scatter at the travelling wave, leading to a broadening of their energy spectra. The ponderomotive potential an electron experiences in this case has the form [16]:

$$U_p \cong 2|E_0|^2 \frac{e^2}{m_0(\omega_1+\omega_2)^2} \cos\left[(\omega_1-\omega_2)t - (\omega_1\cos\alpha - \omega_2\cos\beta)\frac{z}{c} - (\omega_1\sin\alpha - \omega_2\sin\beta)\frac{y}{c}\right],$$

(1)

where $z$ the electron propagation direction. In the electron's rest frame, a standing optical wave is formed. Therefore the electron experiences a constant phase with respect to the light intensity modulation and is pushed out of the high-intensity regions by the ponderomotive force (shown schematically in Figure 1**e**). An arbitrary choice of incident angles $\alpha$ and $\beta$ leads to an angular tilt of the travelling wave with respect to $z$ (term $-(\omega_1\sin\alpha - \omega_2\sin\beta)y/c$ in Eq. (1)) and consequently to the modulation of the electron's transverse momentum. Here we show that for a particular combination of light frequencies and incident angles of the two laser beams, only the longitudinal component of the electron momentum changes.

The laser pulses are obtained by optical parametric generation, where two photons with energies $\hbar\omega_1 > \hbar\omega_2$ and $\hbar\omega_2$ are produced from the incident photon with energy $\hbar\omega = \hbar\omega_1 + \hbar\omega_2$. Diagrams describing the individual scattering events of electrons in vacuum are shown in Figure 1**a-c**. While Figure 1**a** shows the situation before, Figure 1**b** shows the system after the stimulated Compton scattering process where a photon with higher energy is absorbed by an electron while a photon with lower energy is emitted. The incident angles $\alpha$ and $\beta$ of the laser beams are selected in such a way that the electron momentum change $\Delta\hbar\mathbf{k}$ is parallel to its initial momentum $\hbar\mathbf{k}_{in}$, leading to zero transverse momentum change. The second possible process (Figure 1**c**) leads to a decrease of electron energy/momentum.

The relativistic energy and momentum conservation laws for the case of zero transverse momentum transfer $\Delta\hbar\mathbf{k}_\perp = 0$ for the system of an electron and two photons with different energies can be written as:

$$\Delta E_{kin} = (\gamma_f - 1) m_0 c^2 - (\gamma_i - 1) m_0 c^2 = \hbar\omega_1 - \hbar\omega_2,$$

$$\Delta \hbar \mathbf{k}_\perp = 0 = \frac{\hbar\omega_1}{c}\sin\alpha - \frac{\hbar\omega_2}{c}\sin\beta \quad \rightarrow \quad \sin\alpha = \frac{\omega_2}{\omega_1}\sin\beta, \quad (2)$$

$$\Delta \hbar \mathbf{k}_\| = \gamma_f m_0 |\mathbf{v}_f| - \gamma_i m_0 |\mathbf{v}_i| = \frac{\hbar\omega_1}{c}\cos\alpha - \frac{\hbar\omega_2}{c}\cos\beta.$$

Here $\Delta\hbar\mathbf{k}_\|$ is the longitudinal momentum change of the electron, $\Delta E_{kin}$ is the change of the electron kinetic energy, $c$ is speed of light and $\gamma_{f,i} = (1 - |\mathbf{v}_{f,i}|^2/c^2)^{-1/2}$ are the relativistic Lorentz factors corresponding to the final and initial electron velocities $\mathbf{v}_f$ and $\mathbf{v}_i$, respectively. Because the frequencies of the two interacting photons are not independent in our experiment due to the parametric generation process, the set of equations (2) leads to a formula defining the angles $\alpha$ and $\beta$ as a function of the initial electron velocity $\mathbf{v}_i$ and frequency $\omega_1$:

$$\alpha = \arcsin\left[\frac{(\omega - \omega_1)}{\omega_1}\sin\beta\right],$$

$$\beta = \arccos\left[\frac{\hbar^2\omega(2\omega_1 - \omega) - c^2(a - \gamma_i m_0 |\mathbf{v}_i|)^2}{2\hbar c(\omega - \omega_1)(a - \gamma_i m_0 |\mathbf{v}_i|)}\right], \quad (3)$$

$$a = m_0 c\sqrt{b^2 - 1}, \quad b = \frac{2\hbar\omega_1 - \hbar\omega}{m_0 c^2} + \gamma_i.$$

This equation shows that the experimental parameters for purely longitudinal momentum transfer, namely the values of $\alpha$, $\beta$, $\omega_1$ and $\omega_2$, can be adapted to any initial electron velocity $0 \le |\mathbf{v}_i| < c$ (see Supplementary Figure 1 for details).

The experimental demonstration of the inelastic ponderomotive scattering of electrons at a high-intensity optical travelling wave is carried out in a vacuum chamber of a modified scanning electron microscope (SEM), which serves as a pulsed source of electrons that are photoemitted from the SEM cathode (Schottky tip) by an ultraviolet laser pulse. After acceleration by electrostatic fields to the final kinetic energy of $E_{kin}$=29 keV, electrons are focused to the interaction region. Here they experience the optical fields of the two femtosecond laser pulses at wavelengths of $\lambda_1 = 2\pi c/\omega_1 = 1356$ nm ($\hbar\omega_1$=0.91 eV) and $\lambda_2$=1958 nm ($\hbar\omega_2$=0.63 eV). The two beams intersect with the electron beam under angles $α=(41±2)°$ and $β=(107±2)°$ obtained from equation (3). Electron spectra after the interaction are measured by an electromagnetic spectrometer and a micro-channel plate (MCP) detector (Figure 1**d**). For a more detailed description of the experimental setup and its characterization, see Methods.

The measured electron spectra in the presence (red curve) and absence (grey curve) of the optical travelling wave are plotted in Figure 2**a**, in comparison with numerical calculations (dashed curve). With laser pulses in both beams having equal pulse energies of $E_p=85$ μJ, leading to a peak optical intensity of $I_p=3\times10^{15}$ W/cm$^2$, we observe broad shoulders in the spectrum. The large energy modulation of more than 10 keV is a consequence of the high intensity gradient caused by the small period $\lambda_g = 2\pi c/(\omega_1 \cos\alpha - \omega_2 \cos\beta) = 1.41$ μm of the travelling wave.

Due to the periodicity of the scattering potential in the longitudinal direction of the electron wavepacket propagation, coherent interference peaks, separated by the difference between the energies of the two photons participating in the scattering $ΔE=ℏω_1-ℏω_2$, are expected in the electron spectra as a consequence of the quantum interference between electron matter waves scattered by subsequent periods of the travelling wave. This is analogous to the classical Kapitza-Dirac experiment in the diffraction regime [5] with the roles of the transverse and the longitudinal directions exchanged. To reach the coherent regime of the interaction, the longitudinal coherence length of the electron beam has to be significantly longer than $\lambda_g$. This is, however, not the case in our experiment, where the coherence length can be estimated to be $\xi_\parallel=2.35 v_i ℏ/\delta E_{kin,in} \approx 1.8$ μm [18], where $\delta E_{kin,in}=0.5$ eV is the expected initial energy spread (FWHM) of the electron beam.

The maximum observed energy modulation of the electrons corresponds to the simultaneous absorption and emission of more than $10^4$ photons by a single electron. In this high intensity regime, where the scattering rate exceeds the optical frequency of the driving light [4], a depletion of the electron population around the initial energy and the appearance of two broad incoherent rainbow peaks are expected in the electron spectra, similar to the high-intensity Kapitza-Dirac effect [4]. However, the shape of the measured spectra is further influenced by the fact that the electron pulse duration $\tau_{e,FWHM}$ used in the experiment is longer than the durations of the laser pulses $\tau_{1,FWHM}$, $\tau_{2,FWHM}$, generating the optical travelling wave (see Methods for details). The non-ideal temporal overlap leads to a broad distribution of interaction strengths experienced by the electrons. The comparison between calculated electron spectra for the two cases $\tau_{e,FWHM} \gg \tau_{1,FWHM}, \tau_{2,FWHM}$ and $\tau_{e,FWHM} \ll \tau_{1,FWHM}, \tau_{2,FWHM}$ is shown in Supplementary Figure 2.

Due to the incoherent scattering nature of the observed effect, the interaction can be described classically as a scattering of point-like particles at the travelling optical wave. Each calculated spectrum in Figure 2**a**, **b** is therefore obtained by a Monte-Carlo simulation of the interaction between a set of particles and the two laser fields. The interaction is modelled by a numerical integration of the classical relativistic equation of motion with the Lorentz force (see Methods for details).

For further characterization, we measure electron spectra as a function of the energy of the two driving laser pulses in the pulse energy range of 19-128 µJ (solid curves in Figure 2**b**). Again, numerical simulations fit the data very well. The dependence of the induced energy spread $\delta E$ as a function of pulse energy is shown in Figure 2**c**. The spectrum with the maximum observed energy spread of 19.6 keV is obtained with a peak intensity of $I_p = 5 \times 10^{15}$ W/cm$^2$, corresponding to a normalized field amplitude $a_0 = eE_0/(m_0 c\omega) = 0.1 \ll 1$. The experiment presented here thus occurs in the sub-relativistic field regime (for details see Methods).

To characterize the transverse momentum transfer and prove that it is negligible, we measure the 2D spatial distribution of the electrons on the MCP detector after the spectrometer (Figure 2**e**). A modulation of the transverse momentum of electrons would lead to their deflection in *y* and would appear here as a spread along $y_{det}$ axis ($x_{det}$ and $y_{det}$ being the Cartesian coordinates in the detector plane, see Figure 1**d**). The observed deflection angles of electrons are below the angular resolution of the setup of $\delta\theta_{def}\approx20$ mrad, confirming that the transverse momentum transfer is negligibly small compared to the longitudinal momentum change (see Methods).

For applications in particle acceleration by laser fields [19,20], an important property of the demonstrated inelastic scattering is the peak value of the energy gain per unit length, the acceleration gradient. In this proof-of-concept experiment we reach $G_p=dE_{kin}/dz=2.2\pm0.2$ GeV/m (Figure 2**d**). Albeit obtained in a second order ponderomotive process, this gradient is already much higher than typical values reached in state-of-the-art radio-frequency accelerators (~50 MeV/m). In addition it is almost on par with the $G_p=3$ GeV/m obtained using vacuum acceleration of electrons by the longitudinal field of radially polarized few-cycle pulses in the relativistic field regime ($a_0=5$) [21]. The high efficiency makes the ponderomotive interaction between electrons and an optical travelling wave interesting for applications in various particle acceleration schemes, such as in laser-wakefield or dielectric laser acceleration, where it could serve for initial pre-acceleration. Furthermore, because of the independence of the ponderomotive force on the sign of the charge, space-time compression of plasmas is possible.

As an important consequence of the periodic sinusoidal modulation of the longitudinal electron momentum on the femtosecond time scale, an attosecond bunch train is generated due to a rotation of the electron distribution in longitudinal phase space during the ballistic propagation (propagation without any external forces) after the interaction [6,16,17]. This opens a way to reach sub-optical cycle, i.e. attosecond (1 as=$10^{-18}$ s), temporal resolution in ultrafast electron diffraction and microscopy experiments, or to control the electron injection in novel photonics-based accelerators on attosecond time scales [22,23]. Numerical simulation results of the ballistic bunching of electrons after their interaction with the optical travelling wave are shown in Figure 3 for the two central periods of the temporal intensity envelope. The calculation is performed with the experimental parameters used in this study, namely pulse energies of $E_p$=19 µJ (the lowest spectrum in Figure 2**b**) and an initial electron energy spread of $\delta E_{kin,in}$=0.5 eV. The temporal focus occurs already at a propagation distance of ~11 µm after the center of the interaction defined by the point of the intersection of the three beams (the electron beam and the two laser beams) used in the experiment. This corresponds to the propagation time of only 110 fs. The resulting minimum temporal duration of an individual electron bunch in the train is simulated to be $\tau_{FWHM}$=210±10 as, and 30% of all electrons spread initially over one period of the travelling wave ($T_g$=2π/($\omega_1$-$\omega_2$)=14.7 fs) are confined within a temporal window of 300 as in the temporal focus. This number can be further improved by a multistage compression scheme, where, in the first stage, the electron packet will be pre-bunched to a duration of a few femtoseconds by the interaction with a ponderomotive potential of a higher-order Laguerre-Gaussian spatial mode of a focused laser beam [16]. After the first compression stage, the electrons will be injected to a fraction of the period of an optical travelling wave, where the ponderomotive potential can be considered as parabolic in the longitudinal direction. Such a double-stage compression scheme will allow generation of an isolated electron attosecond pulse.

For bunching over macroscopic distances of 100 µm to 5 mm, which would allow temporal compression of the electrons while propagating to a sample under study in ultrafast electron diffraction experiments [24,25], laser pulses with a pulse energy of just 100 nJ to 1 µJ are required. These are readily achievable even for MHz repetition rate laser systems. Interestingly, the attosecond bunch train can be synchronized with an optical pulse produced by difference frequency mixing of the two waves used for the interaction as the optical period of such a pulse matches the time period of the optical travelling wave. In addition and even without carrier-envelope phase stable laser pulses, a passive phase stability is obtained between the difference frequency wave and the optical travelling wave.

The demonstration of the inelastic ponderomotive scattering of electrons at an optical travelling wave presented here further opens the possibilities to study the quantum nature of the nonlinear two-photon inelastic scattering/diffraction processes in vacuum in the energy domain. Similar to single-photon transitions induced by optical near-fields [26-28], two-photon quantum transitions can be observed employing a high resolution electron energy loss spectrometer (EELS) in a transmission electron microscope-based setup. Likewise Ramsey-type interferometry experiments [29,30] are possible via two subsequent interactions. This technique can further serve for energy modulation and bunching of propagating atoms or molecules, where the interaction strength can be even further enhanced using near-resonant interactions [2]. Last, by controlling the polarization state of the two pulses, namely using a combination of linearly and circularly polarized light, energy-resolved studies of electron spin flipping or spin polarization-dependent splitting in high-intensity laser fields might be performed using the presented scheme [31-33].

**Methods**

**Laser pulses**

We use a Ti:sapphire regenerative amplifier (repetition rate $f_{rep}$=1 kHz, pulse duration $\tau_{FWHM}$=90 fs, central wavelength $\lambda$=800 nm, pulse energy $E_p$=7 mJ) to serve as a pump for an optical parametric amplifier (OPA). Here the three laser beams for the experiment are generated. The optical travelling wave is formed by the signal and idler pulses from the OPA with linear polarizations perpendicular to the plane of incidence. The incident angles of the two beams with respect to the electron beam are $\alpha$=(41±2)° and $\beta$=(107±2)°. The spectrum of the two pulses is measured by an infrared spectrometer (see Supplementary Figure 3**a**). Here the central wavelengths of $\lambda_1$=1356 nm ($v_1=\omega_1/2\pi$=221.1 THz), $\lambda_2$=1958 nm ($v_2$=153.1 THz) and the spectral bandwidths of $\delta\lambda_{1,FWHM}$=47 nm ($\delta v_1$=7.66 THz), $\delta\lambda_{2,FWHM}$=86 nm ($\delta v_2$=6.72 THz) are obtained. The time-bandwidth product is measured for a particular OPA output wavelength by frequency-resolved optical gating to be $\tau\Delta v$=0.38 (see Supplementary Figure 3**b**). The pulse durations $\tau_{1,FWHM}$=49±5 fs and $\tau_{2,FWHM}$=56±6 fs are calculated from the spectral widths using the measured value of $\tau\Delta v$. The laser pulses are delivered to the SEM vacuum chamber and focused by two aspherical lenses with the same focal distance $f$=25 mm. The transverse intensity profile of both laser beams is measured by an imaging setup with a charge-coupled device (CCD) camera utilizing the two-photon absorption process. Here the $1/e^2$ radii of $w_{0,1}$=10±0.5 μm and $w_{0,2}$=11.8±0.6 μm are obtained (see Supplementary Figure 3**c**, **d**). The angles of incidence of the two laser beams with respect to the electron propagation direction are measured in the vacuum SEM chamber. The spatio-temporal overlap of the two pulsed beams is reached by monitoring an optical four-wave mixing signal from a thin film of $Al_2O_3$, which is removed during the measurements. The ultraviolet (UV) pulse at a wavelength of $\lambda_{UV}$=251 nm, which serves for photoemission of electrons in the SEM electron gun, is generated via sum-frequency mixing between a part of the signal beam from the OPA and the basic frequency beam from the amplifier, and subsequent second harmonic generation. The UV pulse is focused to the Schottky tip in the SEM electron gun by an UV achromatic lens with a focal length of $f$=15 cm with polarization parallel to the tip axis. The relative timing of all three pulses is controlled using two independent optical delay lines (see the detailed layout of the experimental setup in Supplementary Figure 4).

The normalized field amplitude $a_0$ is typically used to compare the strength of the interaction between laser fields and charged particles. In the relativistic field regime ($a_0 \gg 1$), the field of the laser is strong enough to accelerate the particle close to the speed of light $c$ during one optical cycle. In the opposite case ($a_0 \ll 1$, this study), the change in the electron´s velocity during one optical cycle is small in comparison with $c$ and electrons do not reach relativistic energies (see the simulated energy increase of an accelerated electron during the interaction in Figure 2**d**).

**Electron beam**

The electrons are photoemitted by a single-photon process from a Schottky-type cathode of the SEM using side-illumination. The initial electron energy in the experiment is $E_{kin}$=29 keV. After the electron beam is focused by an objective lens, its transverse size at the interaction point is measured by the knife-edge technique (see Supplementary Figure 5**a**) to be $w$=3.6±0.5 μm ($1/e^2$ radius). The objective lens aperture is removed from the SEM column during the experiments to increase the electron beam current. The electron bunch duration $\tau_{e,FWHM}$=730±30 fs is measured by acquiring the post-interaction electron spectra as a function of the time delay between the UV pulse and the two infrared pulses (see Supplementary Figure 5**b**, **c**).

The transverse momentum of the electrons after the interaction in the plane of incidence of the two laser pulses (*y-z* plane) is characterized in Figure 2**e** by acquiring the 2D image of the electron distribution on the MCP detector. Here any deflection of electrons due to the interaction with the optical travelling wave would lead to spread/tilt of the electron distribution in *y* direction, corresponding to $y_{det}$ coordinate in the detector plane. The observed maximum deflection angle of electrons is below the angular resolution of the setup ~20 mrad. This is limited by defocusing of electrons in $y_{det}$ direction by the edge fields of the magnetic spectrometer (while the electrons are dispersed and focused in $x_{det}$) and the electron beam divergence angle. Negligible observed deflection of interacting electrons agrees with numerical simulations, where we obtain a maximum deflection angle of of $\theta_{def}$≈5 mrad (see Supplementary Figure 6). This is caused by the non-zero width of the angular spectrum of the Gaussian beam plane wave representation in the focus [34], the finite bandwidth of the laser pulses and the fact that the experimental conditions (angles of incidence *α* and *β*, laser frequencies $\omega_1$ and $\omega_2$) based on equation (3) are accurate only for electrons at the initial kinetic energy, while the energy of electrons in the experiment is significantly modulated already during the interaction.

**Detection setup**

Electrons are dispersed by an Elbek-type electromagnetic spectrometer [35] and detected by a Chevron-type MCP detector. The image of the MCP phosphor screen is acquired by a CCD camera. Each spectrum is obtained by integration of the above-threshold signal from 5000 images with an exposure time of 0.1 s. The measured spectra are corrected by the detection efficiency curve of the MCP in the used energy range. The detection setup is calibrated via single-electron counting mode to obtain the number of electrons in each bin for the calculation of the shot noise of the measured electron current. The signal to noise ratio of the normal statistical distribution $SNR = \sqrt{n}$, where $n$ is the detected number of electrons per energy bin, is used for determination of the experimental errors shown in Figure 2**a**, **b**. The dispersion curve of the magnetic spectrometer is calculated to be close to linear for energies $E_{kin}$=20-80 keV. The spectrometer and detection setup performance (dispersion curve, resolution) is verified by a calibration procedure in the range of 20-30 kV by adjusting the accelerating voltage of the SEM. The calibration curve is fitted by a parabola and extrapolated to higher energies. In the experimental spectra, the peak at the electron initial energy is caused by the fact that only ~10% of electrons interact with the laser fields due to the different temporal durations of electron and laser pulses. The measured spectra have a characteristic shape with sharp cut-offs after which the electron count rate decreases approximately linearly with energy. The energy spread $\delta E_{kin}$ is determined by fitting the high and low energy tails of each spectrum with a linear function $f(x)=a\pm bx$. The intersection of the fitted line with the energy axis is considered to be the upper/lower edge ($E_{kin,upper}$, $E_{kin,lower}$) of the particular spectrum and $\delta E_{kin}= E_{kin,upper}- E_{kin,lower}$.

**Simulations**

Simulations are performed by numerical integration of the relativistic equation of motion with the Lorentz force $\frac{d}{dt}(\gamma m_0 \mathbf{v}) = q(\mathbf{E} + \mathbf{v} \times \mathbf{B})$ by a fifth-order Runge-Kutta algorithm. As a result, the time evolution of the electron position $\mathbf{r}(t)$ and the velocity $\mathbf{v}(t)$ are obtained for each particle. The electric and magnetic fields of each pulsed laser beam are considered of Gaussian spatial and temporal mode, in the paraxial approximation:

$$E_x(r,z,t) = E_0 e^{-2\ln 2 \left(\frac{t - \frac{\omega}{c}z}{\tau_{FWHM}}\right)^2} \frac{w_0}{w(z)} e^{-\left(\frac{r}{w(z)}\right)^2} e^{-i\left(kz + \frac{kr^2}{2R(z)} - \varphi(z)\right)}, \quad (4)$$

$$B_y(r,z,t) = \frac{E_x(r,z,t)}{c}, \quad E_y = E_z = B_x = B_z = 0,$$

where $r = \sqrt{x^2 + y^2}$, $k = 2\pi/\lambda$ is the wavevector, $w(z) = w_0\sqrt{1 + (z/z_R)^2}$ the local beam radius, $w_0$ the $1/e^2$ radius of the beam waist, $R(z) = z\left[1 + (z_R/z)^2\right]$ the local radius of curvature of the phase front, $\varphi(z) = \arctan(z/z_R)$ the Gouy phase and $z_R = \pi w_0^2/\lambda$ the Rayleigh length of the beam. Standard rotation transformations are applied to describe the pulses incident under the angles $\alpha$ and $\beta$ in the laboratory coordinate frame. Each electron energy spectrum is calculated in a Monte-Carlo simulation using $10^6$ electrons with a Gaussian distribution in the transverse and longitudinal planes. The final spectra plotted in Figure 2**a**, **b** result from a convolution of the calculated spectra with the response function of the detection setup (grey curve in Figure 2**a**). The initial electron energy spread is assumed to be $\delta E_{kin} = 0.5$ eV (FWHM). Because the experiment is carried out in the regime of <1 electron/bunch, space charge forces are neglected in the simulations. In all simulations, the measured values of the transverse spot sizes and temporal durations of both laser pulses and the electron bunch are used.

**Data availability statement**

The data that support the plots within this paper and other findings of this study are available from the corresponding author upon reasonable request.

**Acknowledgements**

We thank T. Higuchi and J. McNeur for proof-reading of the manuscript. The authors acknowledge funding from ERC grant "Near Field Atto". This publication is funded in part by the Gordon and Betty Moore Foundation through Grant GBMF4744 "Accelerator on a Chip International Program – ACHIP" and by BMBF via a project with contract number 05K16WEC.

**Author contributions**

M. K. planned and carried out the experiments, processed the data and performed the simulations. T. E. designed and fabricated the electromagnetic spectrometer. N. S. developed the controlling software for data acquisition. M. K. and P. H. interpreted the results and wrote the manuscript with contributions of all authors.

**Competing financial interests**

The authors declare no competing financial interests.

**Figures:**

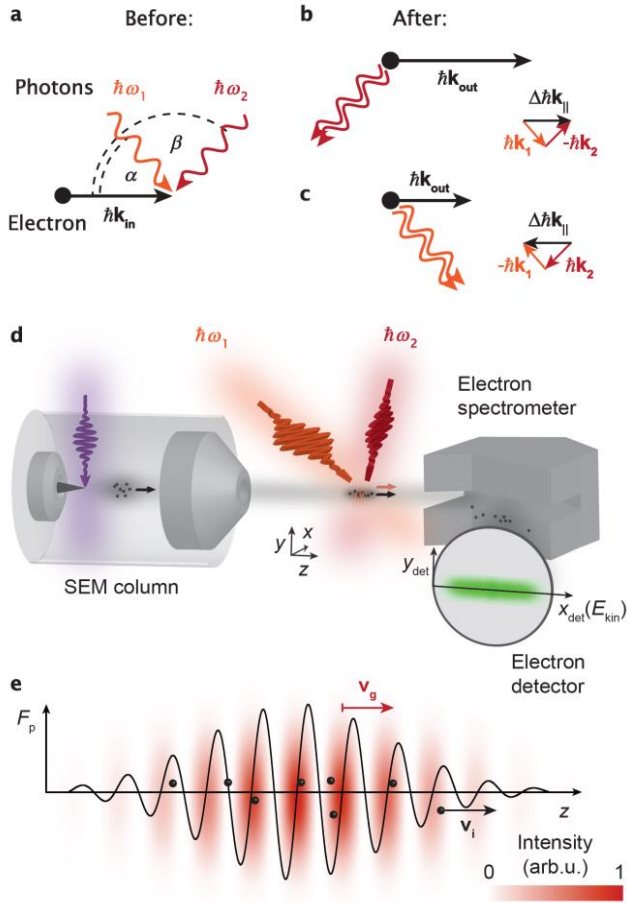

**Figure 1 | Layout of the experiment demonstrating the inelastic ponderomotive scattering of electrons at a high-intensity optical travelling wave in vacuum. a**, Photons with energies $\hbar\omega_1$, $\hbar\omega_2$ ($\hbar\omega_1 > \hbar\omega_2$) and momenta $\hbar\mathbf{k}_1$ and $\hbar\mathbf{k}_2$ intersect with an electron with an initial momentum of $\hbar\mathbf{k}_{in}$ under angles $\alpha$ and $\beta$ (situation before the stimulated Compton scattering). **b**, **c**, The result of an individual stimulated Compton scattering process where a photon with energy $\hbar\omega_1$ is absorbed (emitted) while a photon with energy $\hbar\omega_2$ is emitted (absorbed) leads to an increase (decrease) of the electron energy/longitudinal momentum. **d**, Layout of the experimental setup. Electrons are generated in the electron gun by photoemission using an ultraviolet femtosecond laser pulse. After acceleration and focusing, the electrons interact with the ponderomotive potential of the optical travelling wave created by the two femtosecond laser pulses at frequencies $\omega_1$ and $\omega_2$ in vacuum. The polarization of both pulses is perpendicular to

the plane of incidence. Electron spectra are measured by a magnetic spectrometer and a micro-channel plate detector. **e**, Magnitude of the longitudinal ponderomotive force (black line) of a travelling wave (indicated by red color scale) acting on the co-propagating electrons. The group velocity of the wave $\mathbf{v_g}$ is matched to the initial electron velocity $\mathbf{v_i}$.

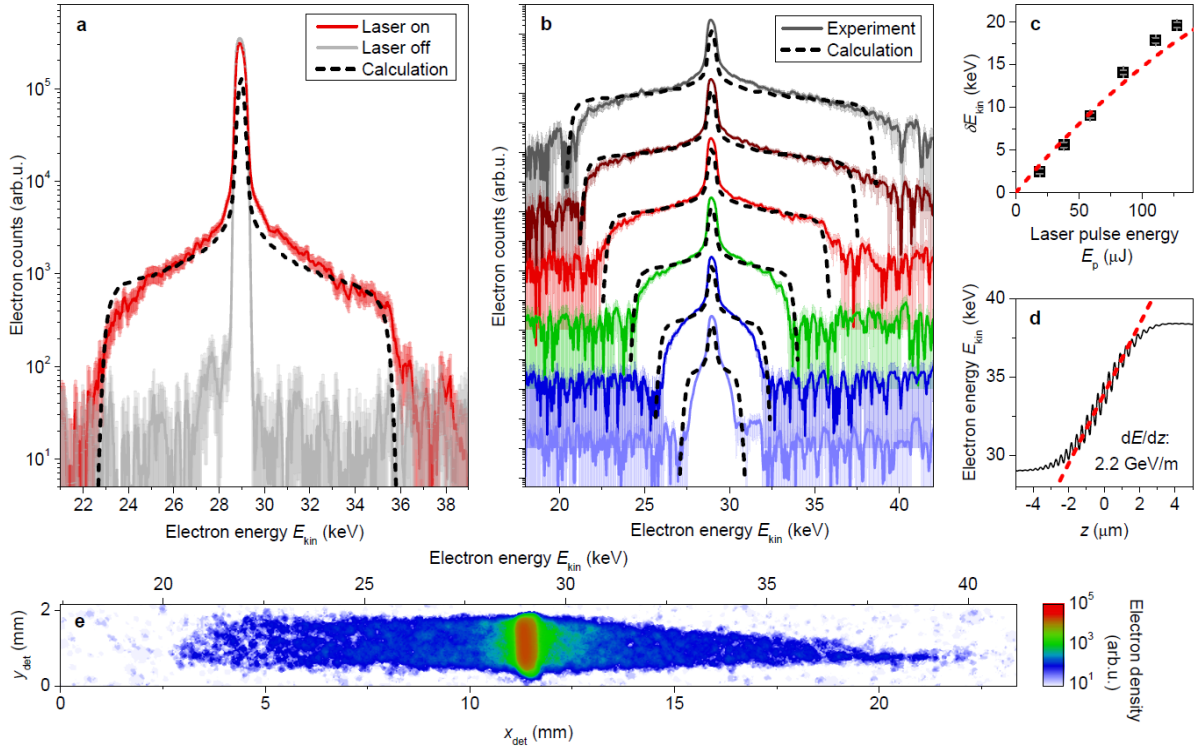

**Figure 2 | Measurements of electron spectra after the inelastic ponderomotive scattering at a high-intensity optical travelling wave in vacuum. a**, Electron spectra in the presence (red curve) and absence (grey curve) of the two laser beams that generate the optical travelling wave, compared to the numerical calculation results (dashed curve). The spectrum with laser on was obtained using pulse energies of 85 μJ in each pulse. **b**, Series of electron spectra obtained with pulse energies of 19 μJ, 38 μJ, 59 μJ, 85 μJ, 111 μJ and 128 μJ (solid curves, bottom to top) with the results of numerical simulations (dashed curves). Spectra are plotted on a logarithmic scale and shifted for clarity. Vertical error bars in panels **a** and **b** (shadowed areas) are calculated as the standard error of the electron count rate detected by the electron-counting micro-channel plate detector (MCP). The error in the energy determination results from the fit of the
19

spectrometer calibration data and equals 0.1 keV (see Methods for details). **c**, Measured (symbols) and calculated (dashed curve) dependence of the final electron energy spread $\delta E_{kin}$ as a function of energy of each of the two laser pulses (for a detailed description of the determination of $\delta E_{kin}$ see Methods). Error bars correspond to the precision of the determination of the electron energy spread from the spectra shown in **b**. **d**, Simulated dependence of the kinetic energy of the electrons with the maximum energy gain on the longitudinal coordinate $z$ (black curve). The slope of the optical cycle-averaged curve directly yields the peak accelerating gradient of $G_p=2.2\pm0.1$ GeV/m. **e**, Image of the electron density detected by the MCP showing the 2D transverse spatial distribution of the electrons (color scale) dispersed by the electromagnetic spectrometer after the interaction with the optical travelling wave.

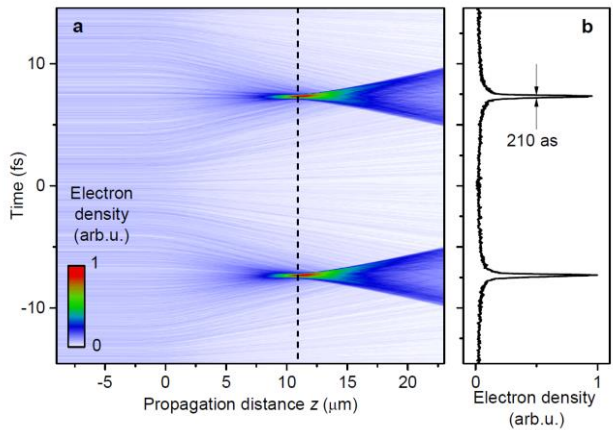

**Figure 3 | Attosecond ballistic bunching of electrons after the interaction with the optical travelling wave. a**, Calculated time evolution of the electron density (color coded) integrated over the transverse plane as a function of the propagation distance $z$ after the interaction with the travelling wave. Due to the induced sinusoidal velocity modulation, the electrons form a series of attosecond bunches after ballistic propagation of 11 μm from the center of the interaction region. The calculation was performed with the parameters used in the experiment, in particular with pulse energies of $E_p=19$ μJ. **b**, Electron density vs. time in the temporal focus (dashed line in **a**) showing bunches with a duration of $\tau_{FWHM}=210\pm10$ as.